\documentclass[pra,superscriptaddress,amsmath,amssymb,twocolumn]{revtex4-2}\newif\ifprep\prepfalse
\usepackage[sort&compress]{natbib}
\usepackage{graphicx}
\usepackage{array}
\usepackage{longtable}%
\usepackage{dcolumn}
\usepackage{bm}
\usepackage{hyperref}

\newcommand{\TabNi}{%
\begin{table*}[ttt]
\caption{\label{tab:Ni25}Fit parameters (cf.\ Equation \ref{eq:PPRCfit}) for the Ni$^{25+}$ DR plasma rate coefficient.  Units are cm$^3$~s$^{-1}$~K$^{3/2}$ for $c_i$ and K for $E_i$. The fit is valid in the temperature range 4600~K to $10^8$~K. In this range the systematic uncertainty of the absolute rate coefficient is 20\% at 90\% confidence level. This set of parameters is slightly different from the one published in Ref.\ \citep{Schippers2000b} since in this earlier work a different extrapolation procedure to high-$n$ resonances was used.}
\begin{ruledtabular}
\begin{tabular}{lrrrrrrr}
 $i$	& \multicolumn{1}{c}{1} & \multicolumn{1}{c}{2} & \multicolumn{1}{c}{3} &  \multicolumn{1}{c}{4} & \multicolumn{1}{c}{5} &	 \multicolumn{1}{c}{6} & \multicolumn{1}{c}{7}\\
 \hline\rule[0mm]{0mm}{4mm}
 $c_i$	& 4.395E$-$2 & 2.715E$-$2 & 1.140E$-$2 & 7.172E$-$3 & 3.070E$-$3 & 2.089E$-$7 & 1.844E$-$9\\
 $E_i$	& 8.133E$+$5 & 4.360E$+$5 & 1.895E$+$5 & 5.758E$+$4 & 3.155E$+$4 & 1.193E$+$4 &	 1.540E$+$2\\
\end{tabular}
\end{ruledtabular}
\end{table*}
}

\newcommand{\TabFeMshell}{%
\ifprep\begingroup\squeezetable\fi
\begin{table*}[ttt]
\caption{\label{tab:FeMshell}Fit parameters (cf.\ Eq.\ \ref{eq:PPRCfit}) for the DR plasma rate coefficient of the Fe$^{q+}$ M-shell ions with $7\leq q \leq 10$ and $13\leq q \leq 15$. Each fit is valid in the specified temperature range $[T_\mathrm{min}, T_\mathrm{max}]$. The quantity $\Delta\alpha$  denotes the systematic uncertainty of the absolute rate coefficient at 90\% confidence level. The row labeled $N\to N'$ specifies which core excitations have been covered by each experiment. Units are cm$^3$~s$^{-1}$~K$^{3/2}$ for $c_i$ and K for $E_i$, $T_\mathrm{min}$, and $T_\mathrm{max}$.}
\begin{ruledtabular}
\begin{tabular}{lrrrrrrr}
 q & \multicolumn{1}{c}{7} & \multicolumn{1}{c}{8} & \multicolumn{1}{c}{9} &\multicolumn{1}{c}{10} &\multicolumn{1}{c}{13} & \multicolumn{1}{c}{14} & \multicolumn{1}{c}{15\footnotemark[5]} \\
 \hline\rule[0mm]{0mm}{4mm}
 $c_1$            & 5.978E$-$7   & 4.777E$-$7   & 6.485E$-$5   & 6.487E$-$5  &   1.570E$-$4 & 1.07E$-$4  & 7.272E$-5$ \\
 $c_2$            & 8.939E$-$7   & 1.231E$-$6   & 6.360E$-$5   & 8.793E$-$5  &   6.669E$-$4 & 8.26E$-$6  & 9.398E$-4$ \\
 $c_3$            & 1.640E$-$5   & 5.055E$-$5   & 3.720E$-$4   & 4.939E$-$4  &   2.789E$-$3 & 1.00E$-$6  & 3.411E$-3$ \\
 $c_4$            & 9.598E$-$5   & 3.413E$-$4   & 1.607E$-$3   & 3.787E$-$3  &   9.938E$-$3 & 1.46E$-$5  & 2.173E$-2$ \\
 $c_5$            & 1.105E$-$4   & 1.625E$-$3   & 3.516E$-$3   & 8.878E$-$3  &   1.362E$-$2 & 2.77E$-$6  & 5.872E$-2$ \\
 $c_6$            & 7.299E$-$4   & 3.873E$-$3   & 7.326E$-$3   & 5.325E$-$2  &   6.888E$-$2 & 1.51E$-$6  & 2.352E$-2$ \\
 $c_7$            & 3.858E$-$3   & 6.438E$-$3   & 2.560E$-$2   & 2.104E$-$1  &   1.838E$-$1 & 3.29E$-$6  & 2.363E$-1$ \\
 $c_8$            & 2.476E$-$2   & 6.970E$-$2   & 1.005E$-$1   &             &              & 1.63E$-$4  & 5.420E$-1$          \\
 $c_9$            & 1.789E$-$1   & 2.925E$-$1   & 1.942E$-$1   &             &              & 4.14E$-$4  & \\
 $c_{10}$         &              &              &              &             &              & 2.17E$-$3  & \\
 $c_{11}$         &              &              &              &             &              & 6.40E$-$3  & \\
 $c_{12}$         &              &              &              &             &              & 4.93E$-$2  & \\
 $c_{13}$         &              &              &              &             &              & 1.51E$-$1  & \\
                  &              &              &              &             &              &            & \\
 $E_1$            & 8.385E$+$0   & 9.034E$+$0   & 3.994E$+$1   & 1.101E$+$2  &   1.088E$+$2 & 7.82E$+$1  & 1.650E$+4$ \\
 $E_2$            & 9.922E$+$1   & 1.128E$+$2   & 5.621E$+$2   & 5.654E$+$2  &   8.388E$+$2 & 1.14E$+$2  & 3.194E$+4$ \\
 $E_3$            & 5.234E$+$2   & 6.624E$+$2   & 1.992E$+$3   & 1.842E$+$3  &   3.006E$+$3 & 2.29E$+$2  & 7.431E$+4$ \\
 $E_4$            & 1.579E$+$3   & 1.143E$+$3   & 8.325E$+$3   & 7.134E$+$3  &   1.127E$+$4 & 2.95E$+$2  & 2.144E$+5$ \\
 $E_5$            & 4.489E$+$3   & 3.926E$+$3   & 2.757E$+$4   & 3.085E$+$4  &   4.162E$+$4 & 5.16E$+$2  & 3.866E$+5$ \\
 $E_6$            & 2.102E$+$4   & 1.300E$+$4   & 7.409E$+$4   & 1.878E$+$5  &   1.885E$+$5 & 7.08E$+$2  & 9.207E$+5$ \\
 $E_7$            & 9.778E$+$4   & 4.684E$+$4   & 1.552E$+$5   & 6.706E$+$5  &   5.422E$+$5 & 1.28E$+$3  & 4.267E$+6$ \\
 $E_8$            & 3.353E$+$5   & 2.670E$+$5   & 4.388E$+$5   &             &              & 2.22E$+$3  & 8.390E$+6$ \\
 $E_9$            & 8.081E$+$5   & 7.358E$+$5   & 7.355E$+$5   &             &              & 3.86E$+$3  & \\
 $E_{10}$         &              &              &              &             &              & 1.12E$+$4  & \\
 $E_{11}$         &              &              &              &             &              & 2.87E$+$4  & \\
 $E_{12}$         &              &              &              &             &              & 1.25E$+$5  & \\
 $E_{13}$         &              &              &              &             &              & 4.45E$+$5  &  \\
                  &              &              &              &             &              &            & \\
 $T_\mathrm{min}$ &       12     &     12       & 100          & 100         &   120        &      11600  & 2300\\
 $T_\mathrm{max}$ &   1.2E$+$9   &   1.2E$+$9   & 1E$+$7       & 1E$+$7      &   1.8E$+$7     &  1.2E$+$8  & 1E$+$14\\
 $\Delta\alpha   $& $\pm25\%$\footnotemark[1]   &  $\pm29\%$\footnotemark[2] &  $\pm25\%$\footnotemark[3]  &  $\pm25\%$\footnotemark[4] &   $\pm29\%$ & $\pm29\%$ & $\pm20\%$ \\
 $N\to N'        $& $3\to3$      &  $3\to3$     & $ 3\to3 $    &  $3\to3$    & $3\to3,3\to4$& $3\to3$    &$ 3\to3, 3\to4$\\
\end{tabular}
\end{ruledtabular}
 \footnotetext[1]{the uncertainty is larger at temperatures below 5800 K, i.e., 38\% at 2300~K and 54\% at 116~K \protect\citep{Schmidt2008a}}
 \footnotetext[2]{the uncertainty is larger at temperatures below 11600 K, i.e., 35\% at 1160~K and 80\% at 116~K \protect\citep{Schmidt2008a}}
 \footnotetext[3]{the uncertainty is larger at temperatures below 10000 K, i.e., 36\% at 1000~K and 82\% at 100~K  \protect\citep{Lestinsky2009}}
 \footnotetext[4]{the uncertainty is larger at temperatures below 1000 K, i.e., 67\% at 330~K and 47\% at 110~K  \protect\citep{Lestinsky2009}}
  \footnotetext[5]{these coefficients supersede the corresponding ones from \href{https://arxiv.org/abs/1002.3678v1}{arXiv.1002.3678v1} and \href{https://www.auburn.edu/academic/cosam/departments/physics/iramp/1_2/index.htm}{Int.\ Rev.\ At.\ Mol.\ Phys.\ {\bf 1}, 109 (2010)}.}
\end{table*}
\ifprep\endgroup\fi
}

\newcommand{\TabFeLshell}{%
\ifprep\begingroup\squeezetable\fi
\begin{table*}[ttt]
\caption{\label{tab:FeLshell}Same as Table II but for  Fe$^{q+}$ L-shell ions with $16\leq q \leq 22$.}
\begin{ruledtabular}
\begin{tabular}{lrrrrrrr}
 q & \multicolumn{1}{c}{16\footnotemark[1]} & \multicolumn{1}{c}{17} & \multicolumn{1}{c}{18} &\multicolumn{1}{c}{19} &\multicolumn{1}{c}{20} & \multicolumn{1}{c}{21} & \multicolumn{1}{c}{22} \\
 \hline\rule[0mm]{0mm}{4mm}
 $c_1$            &    2.06E$-$1   &     4.79E$-$6  &     2.14E$-$5   &     1.24E$-$4  &     7.71E$-$5  &  1.46E$-$4  &  2.47E$-$6\\
 $c_2$            &    1.24E$+$0   &     9.05E$-$5  &     1.05E$-$5   &     1.84E$-$4  &     6.08E$-$5  &  1.17E$-$3  &  1.21E$-$4\\
 $c_3$            &                &     3.48E$-$5  &     4.34E$-$5   &     1.47E$-$4  &     3.93E$-$4  &  4.22E$-$3  &  2.18E$-$3\\
 $c_4$            &                &     1.83E$-$4  &     6.62E$-$5   &     7.87E$-$4  &     1.14E$-$3  &  2.80E$-$3  &  1.50E$-$3\\
 $c_5$            &                &     5.26E$-$4  &     3.86E$-$4   &     3.54E$-$3  &     7.63E$-$3  &  9.22E$-$3  &  1.47E$-$2\\
 $c_6$            &                &     2.12E$-$3  &     1.24E$-$3   &     5.02E$-$3  &     1.37E$-$2  &  3.13E$-$2  &  3.14E$-$2\\
 $c_7$            &                &     4.29E$-$3  &     5.56E$-$3   &     1.96E$-$2  &     2.08E$-$2  &  9.98E$-$2  &  7.98E$-$2\\
 $c_8$            &                &     3.16E$-$2  &     4.07E$-$2   &     6.43E$-$2  &     7.86E$-$2  &                     &  1.25E$-$1\\
 $c_9$            &                &                &     2.92E$-$1   &                &                &             &  7.25E$-$1\\
 $c_{10}$         &                &                &     1.46E$+$0   &                &                &             &           \\
                  &                &                &                 &                &                &             &           \\
 $E_1$            &    4.38E$+$6   &     2.58E$+$3  &     1.01E$+$2   &     1.61E$+$1  &     6.75E$+$1  &  1.58E$+$3  &  1.04E$+$3\\
 $E_2$            &    7.98E$+$6   &     6.08E$+$3  &     2.45E$+$2   &     7.90E$+$1  &     1.15E$+$2  &  2.43E$+$3  &  5.44E$+$3\\
 $E_3$            &                &     1.35E$+$4  &     8.80E$+$2   &     7.73E$+$2  &     2.43E$+$3  &  4.85E$+$3  &  1.35E$+$4\\
 $E_4$            &                &     2.92E$+$4  &     7.53E$+$3   &     3.86E$+$3  &     6.71E$+$3  &  1.09E$+$4  &  3.73E$+$4\\
 $E_5$            &                &     7.62E$+$4  &     1.93E$+$4   &     1.66E$+$4  &     2.74E$+$4  &  7.13E$+$4  &  1.21E$+$5\\
 $E_6$            &                &     2.21E$+$5  &     6.75E$+$4   &     6.08E$+$4  &     6.73E$+$4  &  2.72E$+$5  &  3.15E$+$5\\
 $E_7$            &                &     6.57E$+$5  &     2.90E$+$5   &     2.33E$+$5  &     2.80E$+$5  &  9.92E$+$5  &  9.08E$+$5\\
 $E_8$            &                &     1.40E$+$6  &     1.11E$+$6   &     9.74E$+$5  &     1.11E$+$6  &                     &  4.64E$+$6\\
 $E_9$            &                &                        &     4.65E$+$6   &                        &                        &                     &  1.15E$+$7\\
 $E_{10}$         &                &                &     9.14E$+$6   &                &                &             &           \\
                  &                &                &                 &                &                &             &           \\
 $T_\mathrm{min}$ & 12             &     580        &     120         &     12         &     12         &    230      &   12      \\
 $T_\mathrm{max}$ & 1.2E$+$9       &    1.2E$+$8    &  1.2E$+$8       &  1.2E$+$8      &   1.2E$+$8     &   1.2E$+$8  &  1.2E$+$8 \\
 $\Delta\alpha   $&  $\pm22\%$ &    $\pm20\%$   & $\pm20\%$     &  $\pm20\%$     &  $\pm20\%$ &  $\pm20\%$  & $\pm20\%$ \\
 $N\to N'        $&$   2\to3    $  &$   2\to2    $  & $ 2\to2,2\to3 $ &$  2\to2     $  &  $2\to2$       & $ 2\to2   $ &$ 2\to2, 2\to3$\\
\end{tabular}
 \footnotetext[1]{fit parameters taken from  \protect\citep{Zatsarinny2004b}}
\end{ruledtabular}
\end{table*}
\ifprep\endgroup\fi
}

\newcommand{\FigNi}[1]{%
\begin{figure}[#1]
\includegraphics[width=\columnwidth]{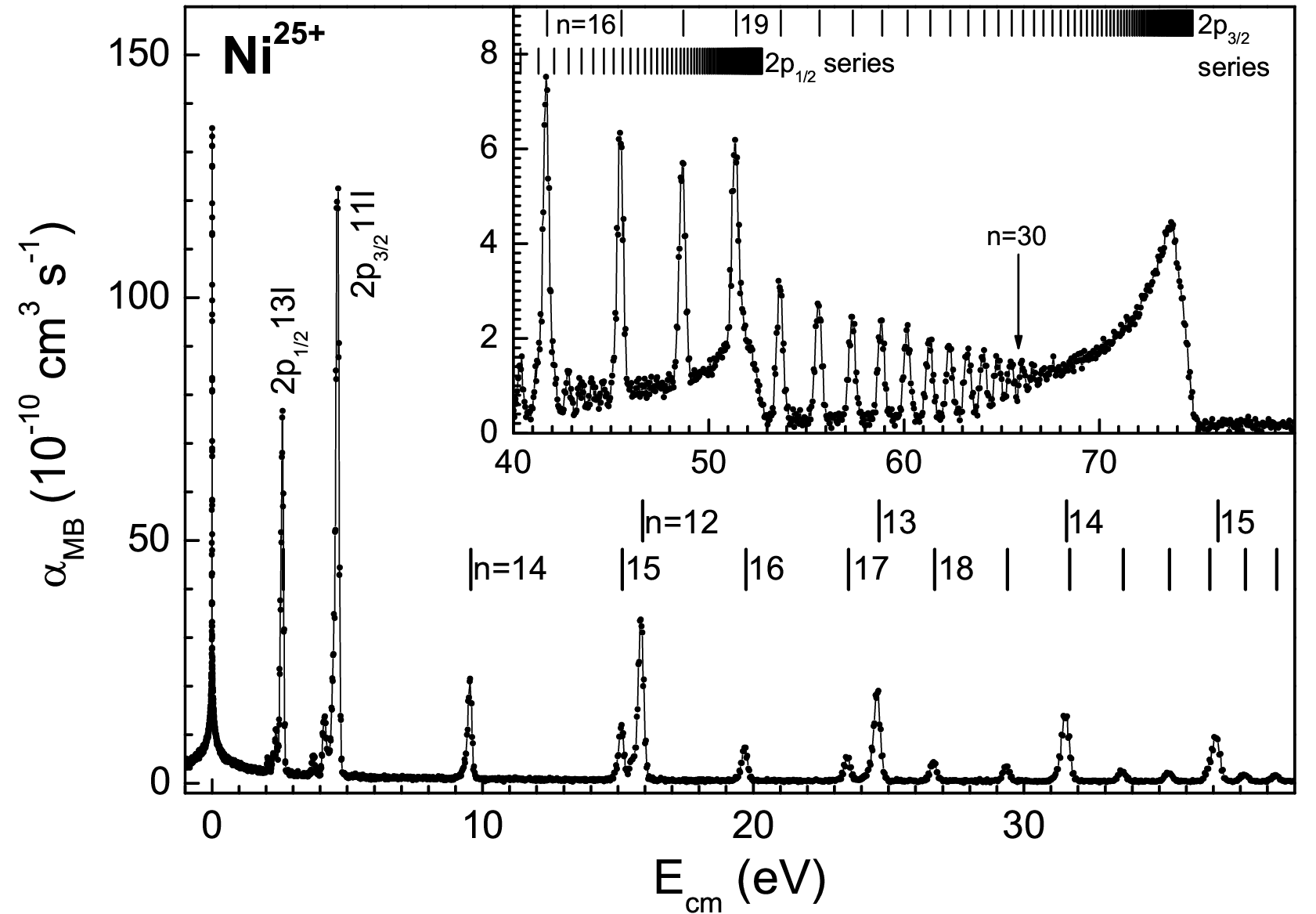}
\caption{\label{fig:Ni25}Measured merged-beams rate coefficient for the recombination of Li-like Ni$^{25+}$ ions with free electrons \citep{Schippers2000b}. Two Rydberg series of  $2p_{1/2}\,nl$ and $2p_{3/2}\,nl$ $\Delta N=0$ DR resonances are discernible converging to the respective series limits at 52.92 eV and 74.96 eV \citep{Ralchenko2008}. The vertical marks denote resonance positions calculated with Equation \ref{eq:Ryd}. The sharp structure at $E_{cm} = 0$ is due to radiative recombination. Positive (negative) energies correspond to electron velocities larger (smaller) than the ion velocity}
\end{figure}
}

\newcommand{\FigNiPlasma}[1]{%
\begin{figure}[#1]
\includegraphics[width=\columnwidth]{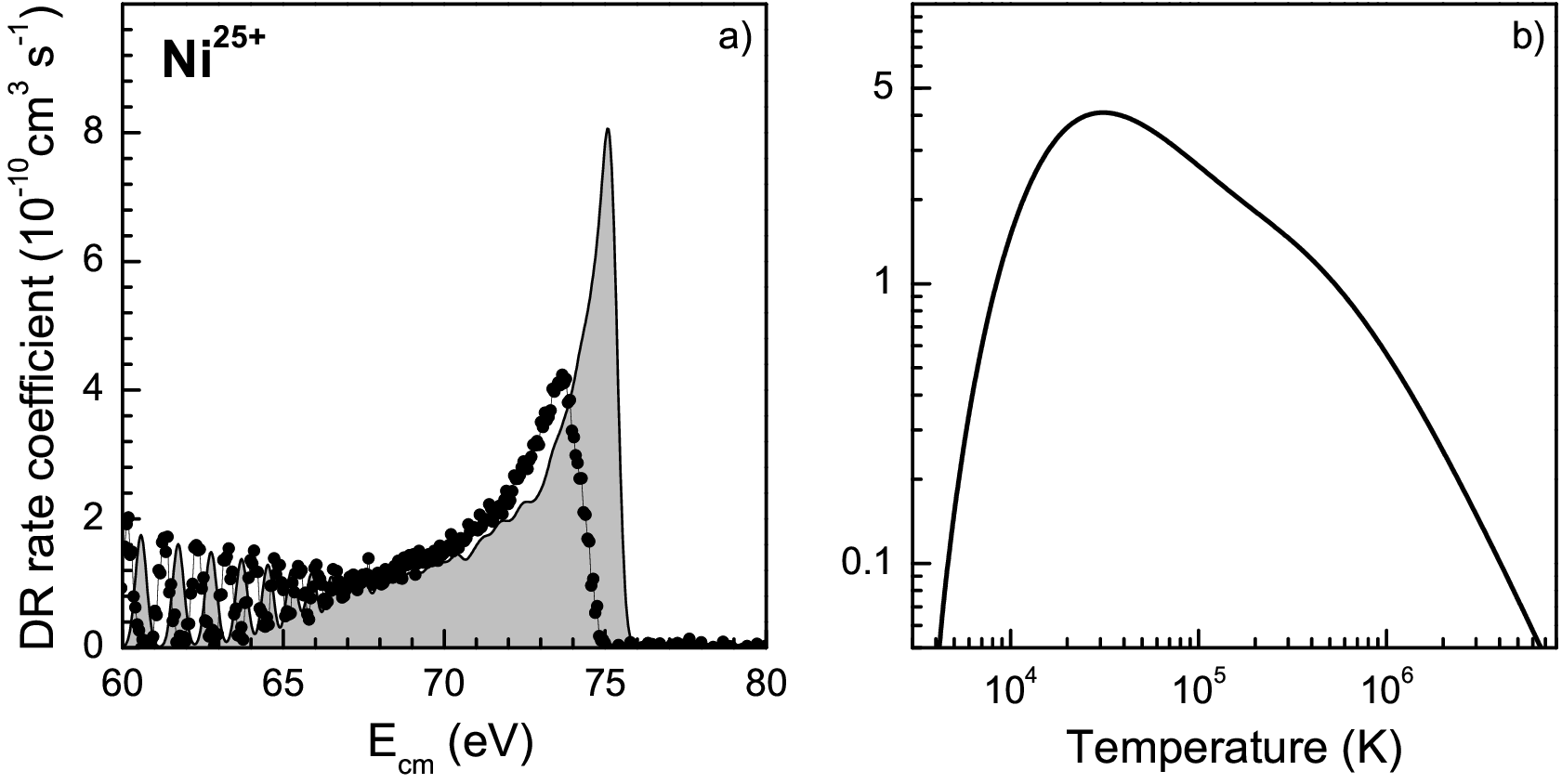}
\caption{\label{fig:Ni25plasma}a) Comparison of the Ni$^{25+}$ merged-beams DR rate coefficient with a theoretical calculation using the AUTOSTRUCTURE code \citep{Badnell1986} at the $2p_{3/2}$ series limit. The theoretical rate coefficient contains contributions by Rydberg resonances beyond the experimental field ionization cut-off at $n=150$. b) Experimentally derived Ni$^{25+}$ $\Delta N=0$ DR plasma rate coefficient obtained by convoluting the ``RR background''-subtracted and  AUTOSTRUCTURE-extrapolated experimental spectrum  with an isotropic Maxwell-Boltzmann electron energy distribution.  The curve was obtained by using Eq.\ \ref{eq:PPRCfit} with the parameters from Table \ref{tab:Ni25}.}
\end{figure}
}

\newcommand{\FigFeSeventeen}[1]{%
\begin{figure}[#1]
\includegraphics[width=\columnwidth]{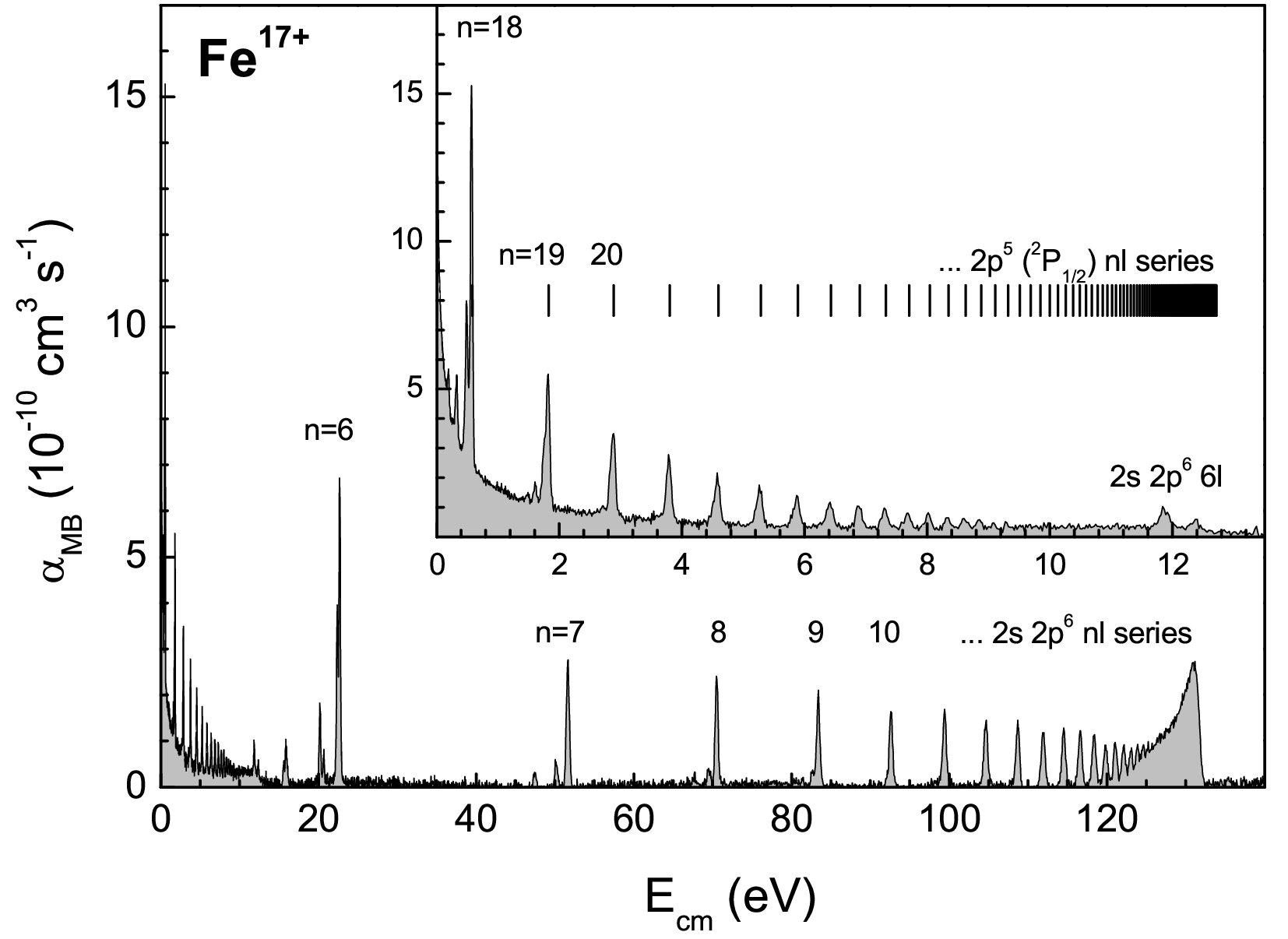}
\caption{\label{fig:Fe17}Measured merged-beams rate coefficient for the recombination of F-like Fe$^{17+}$($2s^2\,2p^5\;^2P_{3/2}$) ions with free electrons \citep{Savin1997,Savin1999}. The inset zooms into the $2s^2\,2p^5\;(^2P_{1/2})\,nl$ Ryd\-berg series of DR resonances associated with $2s^2\,2p^5\;(^2P_{3/2})\to 2s^2\,2p^5\;(^2P_{1/2})$ fine-structure ($\Delta N=0$) core excitations which dominates the DR plasma rate coefficient at low-temperatures.}
\end{figure}
}

\newcommand{\FigFeSeven}[1]{%
\begin{figure}[#1]
\includegraphics[width=\columnwidth]{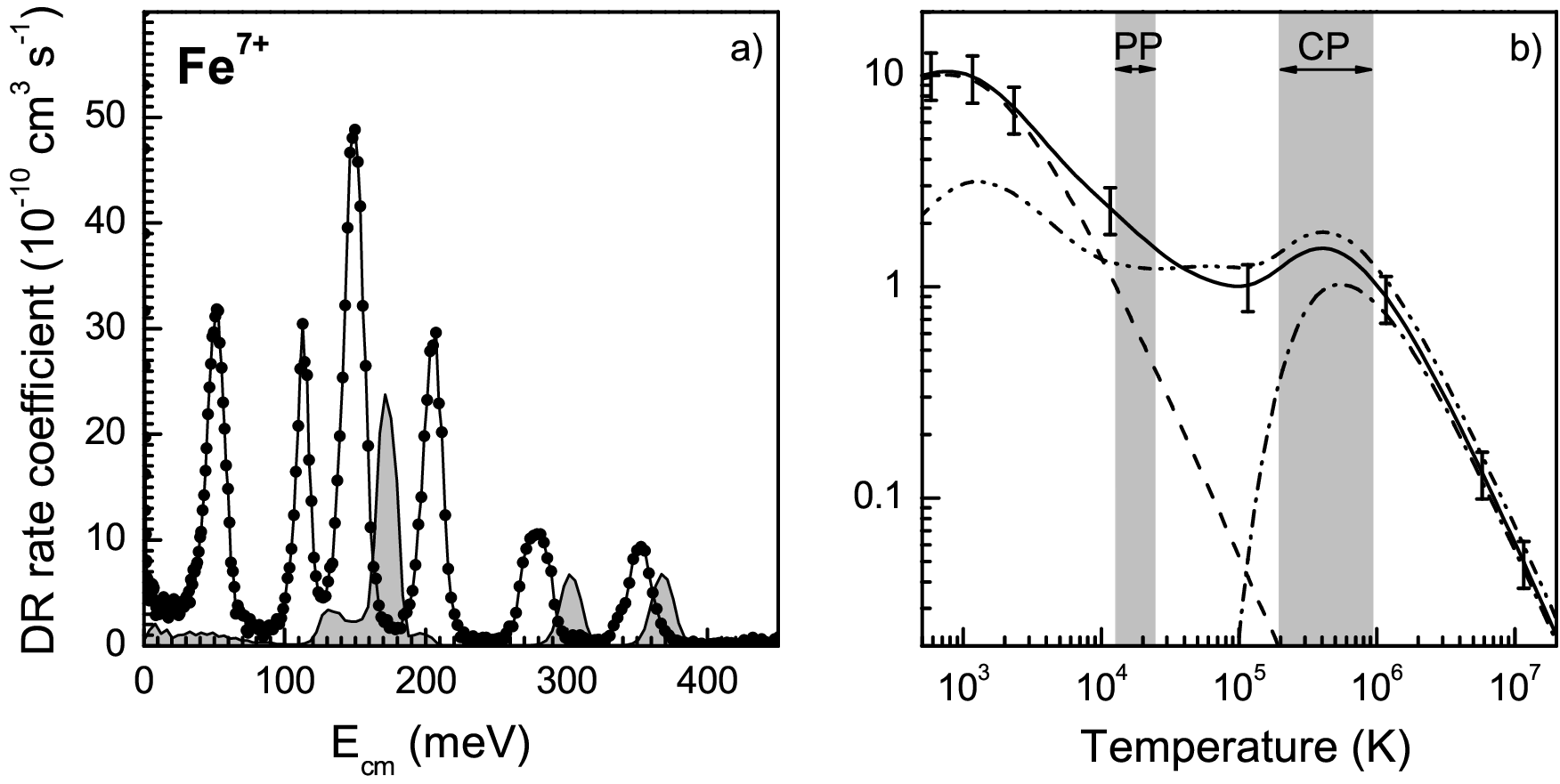}
\caption{\label{fig:Fe7}a) Experimental (filled circles)  and theoretical (shaded curve) merged-beams rate coefficient
for the recombination of electrons with Fe$^{7+}$($3s^3\,3p^6\,3d$) ions at low electron-ion collision energies \citep{Schmidt2008a}. b)  Fe$^{7+}$ DR plasma rate coefficient. The experimental result from the TSR (solid curve with error bars comprising systematic and statistical experimental uncertainties) is compared with the theoretical rate coefficient from the widely used compilation of Arnaud \& Raymond  \cite{Arnaud1992} (dash dotted curve) and with a recent state-of-the-art calculation \citep{Schmidt2008a} (dash-dot dotted curve).  The dashed curve is the contribution to the total plasma rate coefficient of the low-energy DR resonances with $E_\mathrm{res}<0.4$~eV shown in panel a). The temperature ranges where Fe$^{7+}$ is predicted to exist in photoionized plasmas (PP) and collisionally ionized plasmas (CP) are highlighted.}
\end{figure}
}

\newcommand{\FigFeTwentytwo}[1]{%
\begin{figure}[#1]
\includegraphics[width=\columnwidth]{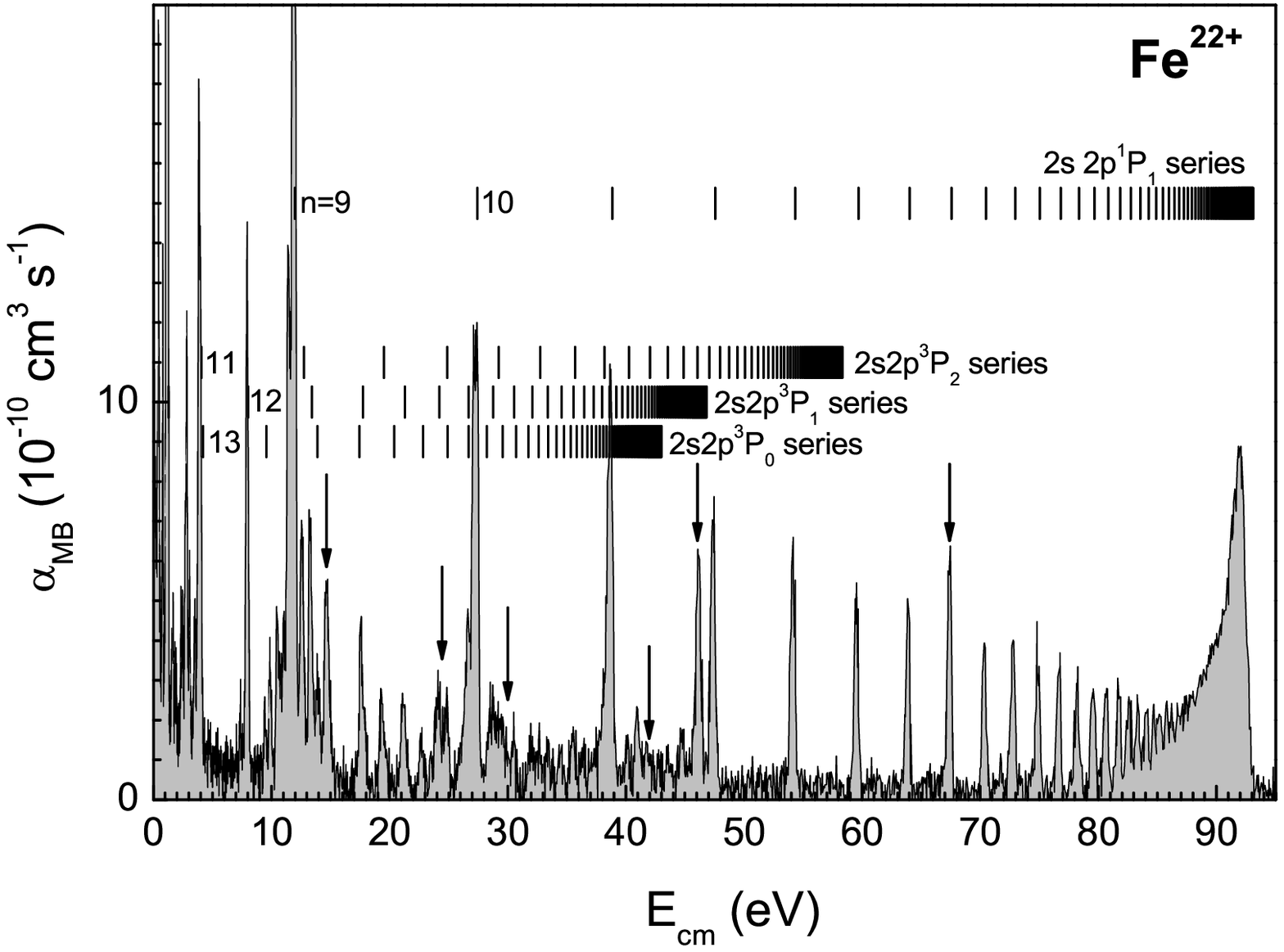}
\caption{\label{fig:Fe22}Measured merged-beams rate coefficient for the recombination of Be-like Fe$^{22+}$ ions with free electrons \citep{Savin2006a}. Vertical arrows mark positions of prominent TR resonances which have been identified by using Eq.\ \ref{eq:Ryd} with $2s^2\to2p^2$ excitation energies from the NIST atomic spectra data base \citep{Ralchenko2008}. In the order of increasing energy these TR resonances are $2p^2\,({^1\!}D_2)\;7l$, $2p^2\,({^3\!}P_1)\;8l$, $2p^2\,({^3\!}P_2)\;8l$, $2p^2\,({^1\!}S_0)\;7l$, a blend of $2p^2\,({^1\!}D_2)\;8l$ and $2p^2\,({^3\!}P_1)\;9l$, and a blend of $2p^2\,({^1\!}D_2)\;9l$ and $2s\,2p\,({^1\!}P_1)\;16l$.}
\end{figure}
}

\newcommand{\FigFeEighteen}[1]{%
\begin{figure}[#1]
\includegraphics[width=\columnwidth]{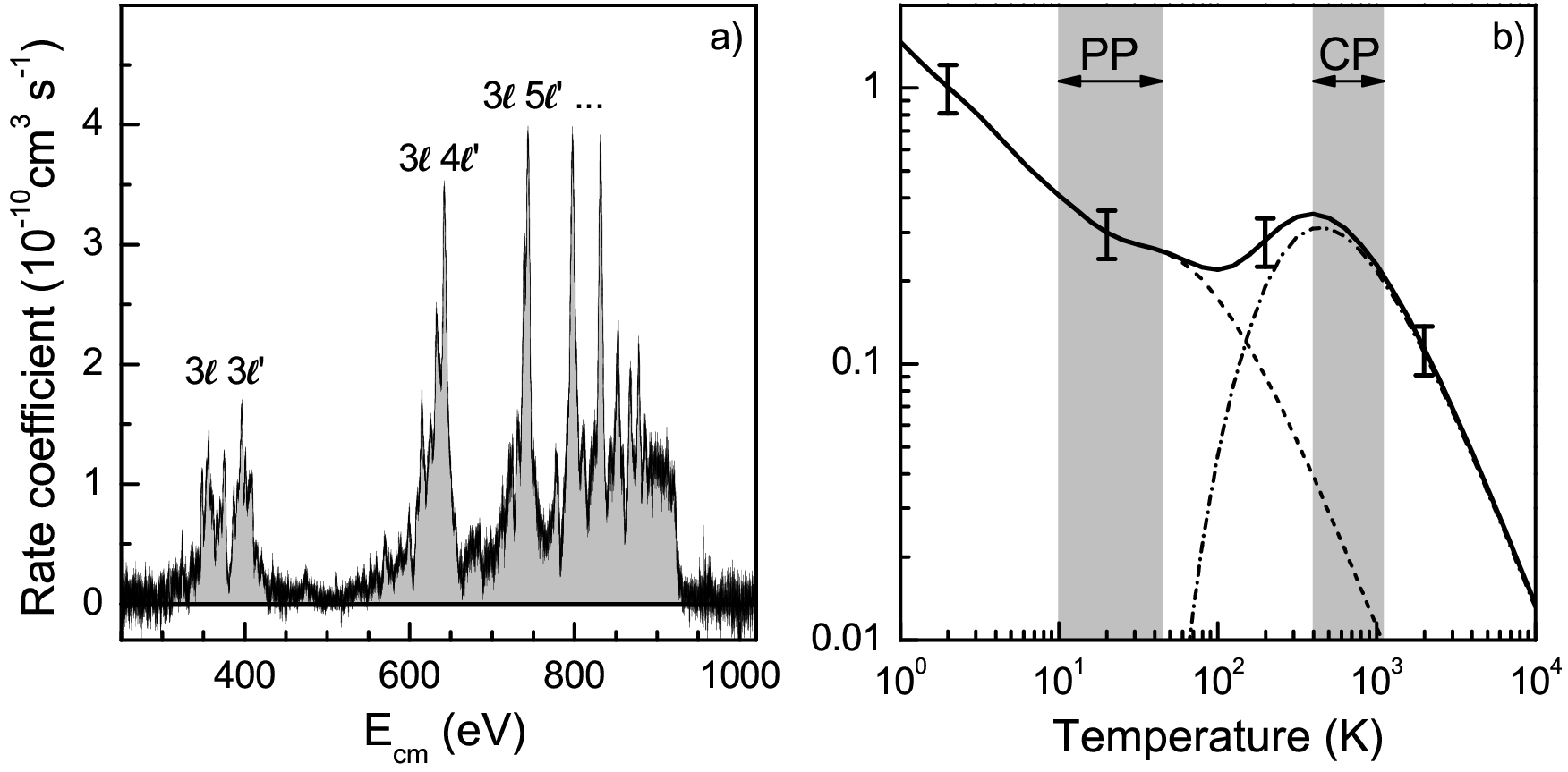}
\caption{\label{fig:Fe18}a) Measured merged-beams rate coefficient for the recombination of O-like Fe$^{18+}$ ions with free electrons \citep{Savin2002c} in the energy range of $N=2\to N'=3\; (\Delta N=1)$ DR resonances.
b) Experimentally derived rate coefficient for DR of Fe$^{18+}$ in a plasma \citep[full line with error bars denoting the combined systematic and statistical experimental uncertainties,][]{Savin2002c}. The dotted and dashed lines are the contributions by $\Delta N=0$ DR and $\Delta N=1$ DR, respectively.
The temperature ranges where Fe$^{18+}$ forms in photoionized plasmas (PP) and collisionally ionized plasmas (CP) are indicated.}
\end{figure}
}

\newcommand{\FigFeMshell}[1]{%
\begin{figure}[#1]
\flushleft{\includegraphics[width=0.96\columnwidth]{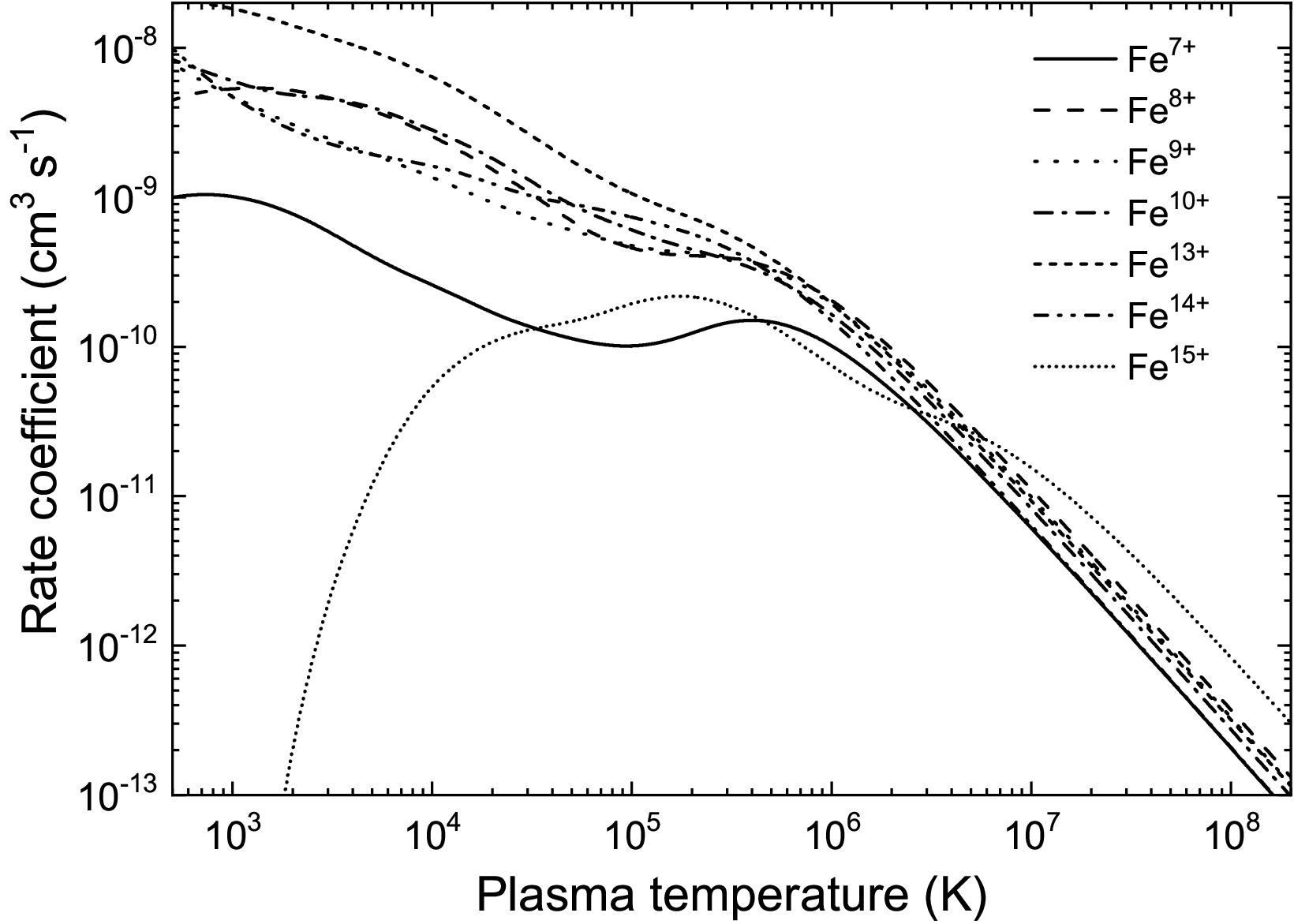}}
\caption{\label{fig:FeMshell}Presently available, experimentally derived rate coefficients for  DR of Fe M-shell ions in a plasma. The curves were obtained by using Eq.\ \ref{eq:PPRCfit} with the  parameters from Tab.\ \ref{tab:FeMshell}}
\end{figure}
}

\newcommand{\FigFeLshell}[1]{%
\begin{figure}[#1]
\includegraphics[width=\columnwidth]{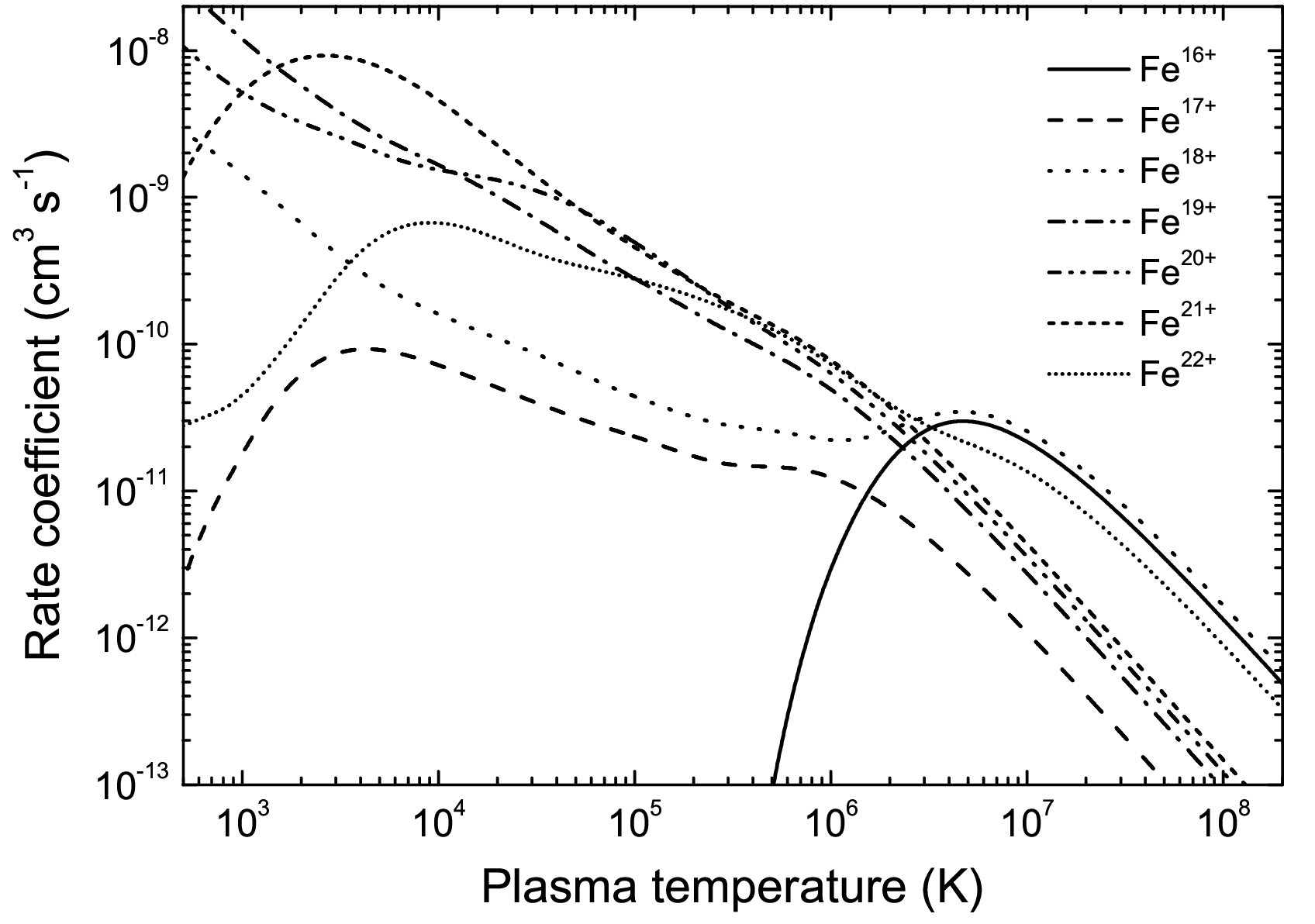}
\caption{\label{fig:FeLshell}Same as Fig.~7 but for Fe L-shell ions. The curves were obtained by using Eq.\ \ref{eq:PPRCfit} with the parameters from Tab.\ \ref{tab:FeLshell}}
\end{figure}
}

\begin{document}

\title{Dielectronic recombination data for astrophysical applications:  Plasma rate-coefficients for  Fe$^\mathbf{q+}$ ($\mathbf{q}$=7--10, 13--22) and Ni$^{25+}$ ions from storage-ring experiments}

 \author{S. Schippers}
 \affiliation{Institut f\"{u}r Atom- und Molek\"{u}lphysik, Justus-Liebig-Universit\"{a}t, Leihgesterner Weg 217, 35392 Giessen, Germany}

 \author{M. Lestinsky}
 \affiliation{Max-Planck-Institut f\"{u}r Kernphysik, Saupfercheckweg 1, 69117 Heidelberg, Germany}
 \affiliation{Columbia Astrophysics Laboratory, Columbia University, 550 West 120th, New York, NY 10027, USA}

 \author{A. M\"{u}ller}
 \affiliation{Institut f\"{u}r Atom- und Molek\"{u}lphysik, Justus-Liebig-Universit\"{a}t, Leihgesterner Weg 217, 35392 Giessen, Germany}

 \author{D. W. Savin}
 \affiliation{Columbia Astrophysics Laboratory, Columbia University, 550 West 120th, New York, NY 10027, USA}

 \author{E. W. Schmidt}
 \affiliation{Institut f\"{u}r Atom- und Molek\"{u}lphysik, Justus-Liebig-Universit\"{a}t, Leihgesterner Weg 217, 35392 Giessen, Germany}

 \author{A. Wolf}
 \affiliation{Max-Planck-Institut f\"{u}r Kernphysik, Saupfercheckweg 1, 69117 Heidelberg, Germany}


\begin{abstract}
This review summarizes the present status of  an ongoing experimental effort to provide reliable rate coefficients for dielectronic recombination of highly charged iron ions for the modeling of astrophysical and other plasmas. The experimental work has been carried out over more than a decade at the heavy-ion storage-ring TSR of the Max-Planck-Institute for Nuclear Physics in Heidelberg, Germany. The experimental and data reduction procedures are outlined.
The role of previously disregarded processes such as fine-structure core excitations and trielectronic recombination is highlighted. Plasma rate coefficients for dielectronic recombination of Fe$^{q+}$ ions (${q}$=7--10, 13--22) and Ni$^{25+}$ are presented graphically and in a simple parameterized form allowing for easy use in plasma modeling codes. It is concluded that storage-ring experiments are presently the only source for reliable low-temperature dielectronic recombination rate-coefficients of complex ions.
\end{abstract}



\maketitle

\section{Introduction}

Dielectronic recombination (DR) is an important electron-ion collision process governing the charge balance in atomic plasmas \citep{Savin2007d,Badnell2007b}. Accurate DR rate coefficients are therefore required --- as well as many other atomic data --- for the interpretation of observations of such plasmas be they man-made or astrophysical. Because of the vast atomic data needs most of the data that are presently used in plasma modeling codes have been generated by theoretical calculations. In order to assess the reliability of these calculations and to point out directions for their improvements benchmarking experiments are vitally needed \citep{Ferland2003a,Kallman2007a}.

For more than a decade, our collaboration has performed measurements of absolute DR rate coefficients employing the electron-ion merged-beams method at the heavy-ion storage ring TSR of the Max-Planck-Institute for Nuclear Physics in Heidelberg, Germany.  Status reports on these activities have been presented repeatedly \citep{Mueller1999c,Schippers1999b, Savin2000c,Mueller2001b,Gwinner2001,Savin2001a,Savin2005a,Schippers2005a,Schwalm2007,Mueller2008a} and  a comprehensive bibliography of storage-ring DR measurements with astrophysically relevant ions has been published recently \citep{Schippers2009a}.

In particular, we have concentrated on iron ions because of their prominent role in X-ray astronomy.
Iron is the most abundant heavy element \citep{Asplund2009} and still contributes to line emission from astrophysical plasmas when lighter elements are already fully stripped. Line emission from iron ions is prominent in many spectra taken with the X-ray observatories such as XMM-Newton and Chandra \citep{Paerels2003a}.

In DR an initially free electron excites another electron, which is initially bound on the primary ion, and thereby looses enough energy such that it becomes bound, too. The DR process is completed if in a second step the intermediate doubly excited state decays radiatively to a state below the ionization threshold of the recombined ion. The initially bound core electron may be excited from a state with principal quantum number $N$ to a state with principal quantum number $N'$. There are an infinite number of excitation channels. In practice, however, only the smallest excitation steps with $N'=N$ ($\Delta N=0$ DR) , $N'=N+1$ ($\Delta N=1$ DR) and sometimes also $N'=N+2$ ($\Delta N=2$ DR) contribute significantly to the total DR rate coefficient. In the measurements reported here only $\Delta N=0$ DR and  $\Delta N=1$ DR have been considered.

Energy conservation dictates that the DR resonance energies $E_\mathrm{res}$ are given by $E_\mathrm{res} = E_d-E_i$ with $E_d$ and $E_i$ being the total electron energies of the doubly excited resonance state and the initial state, respectively. Both $E_d$ and $E_i$ can amount to several 100 keV. In contrast, $E_\mathrm{res}$ can be less than 1~eV. Calculating such a small difference of two large numbers with sufficient accuracy is a considerable challenge even for state-of-the-art atomic structure codes \citep{Badnell2007b}. Unfortunately, small uncertainties in low-energy DR resonance positions
can translate into huge uncertainties of the calculated DR rate coefficient in a plasma \citep{Schippers2004c}. Therefore, experimental DR measurements are particulary valuable for ions with strong resonances at energies of up to a few eV which decisively determine the low-temperature DR rate coefficients important for near neutrals in an electron ionized plasma or for complex ions in photoionized plasmas. Almost all iron ions more complex than helium-like belong to this latter class.

Cosmic atomic plasmas can be divided into collisionally ionized plasmas (CP) and photoionized plasmas (PP) \citep{Savin2007d} both covering broad temperature ranges. Historically, most theoretical recombination data were calculated for CP (see e.g. \citep{Mazzotta1998}) where highly charged ions exist only at rather large temperatures, e.g., in the solar corona. At these temperatures, recombination rate coefficients are largely insensitive to low-energy DR resonances. Consequently, the theoretical uncertainties are much smaller at higher than at lower plasma temperatures which are typical for PP. If the CP rate coefficients are also used for the astrophysical modeling of PP, inconsistencies arise. This has been noted, e.g., in the astrophysical modeling of X-ray spectra from active galactic nuclei by Netzer \citep{Netzer2004a} and Kraemer et al. \citep{Kraemer2004a}.

It is clear, that the large discrepancies at low temperatures between the experimental and the early theoretical rate coefficients are due to a simplified theoretical treatment that was geared towards CP and more or less disregarded low-energy DR in order to keep the calculations tractable. Modern computers allow more sophisticated approaches, and recent theoretical work has aimed at providing a more reliable recombination data-base by using state-of-the-art atomic codes \cite{Badnell2007b}. Badnell and coworkers \citep{Badnell2003a} have calculated DR rate coefficients for finite-density plasmas. Results have been published for the isoelectronic sequences from H-like to Mg-like \citep[and references therein]{Altun2007a}. Rate coefficients for non-resonant radiative recombination (RR) are also available \citep{Badnell2006d,Trzhaskovskaya2009}. Independently, Gu calculated DR and RR rate coefficients for selected ions of astrophysical interest \citep{Gu2003b,Gu2003c,Gu2004a}. Although the new theoretical work removes the striking low-temperature disagreement that was found between experimental and early theoretical results, significant theoretical uncertainties remain as discussed above. Experimental benchmarks are thus indispensable for arriving at a reliable DR data base for the astrophysical modeling, particularly of low-temperature plasmas.

The present review is organized as follows. The experimental procedure for obtaining a DR rate coefficient from a storage-ring experiment is outlined in Section \ref{sec:exp}. In Section \ref{sec:plasma} the various steps for deriving a plasma rate coefficient from the measured data are briefly discussed. In Section \ref{sec:iron} illustrative examples are given for selected iron ions and, finally, fit parameters for a convenient representation of our experimentally derived DR plasma rate coefficients are listed for Ni$^{25+}$ and Fe$^{q+}$ with (${q}$=7--10, 13--22). Because of their dependence on subtle details of the  particular atomic structure of each ion species reliable DR rate coefficients cannot be obtained by interpolations along isonuclear or isoelectronic sequences of ions, i.e., a separate measurement has been carried out for each individual ion. Lithium-like Ni$^{25+}$ has been included in this compilation since its DR resonance structure is relatively simple and therefore serves as a pedagogical example for the presentation of the experimental technique and the subsequent data analysis.
Throughout this paper we refer exclusively to the charge states of the primary ions before recombination.

\section{Experimental procedure}\label{sec:exp}

Heavy-ion storage rings equipped with electron coolers serve as an excellent experimental environment for electron-ion collision studies \citep{Mueller1997c,Schuch2007a}. In electron-ion merged-beams experiments at heavy-ion storage-rings a fast-moving ion beam is collinearly merged with a magnetically guided electron beam with an overlap length $L$ of the order of  1--2 m. Recombined ions are separated from the primary beam in the first bending  magnet downbeam of the interaction region and directed onto a single particle detector.  Since the reaction products are moving fast and are confined in a narrow cone they can easily be detected with an efficiency $\eta$ of nearly 100\%.

\FigNi{bbb}

Storage rings measure the electron-ion (RR+DR) recombination cross section times the relative velocity convolved with the energy spread of the experiment, called a merged beams recombination rate coefficient (MBRRC).  This differs from a plasma recombination rate coefficient (PRRC) for a Maxwellian temperature distribution. From the measured count rate $R$, the stored ion current $I_i$, and the electron density $n_e$ of the electron beam, the MBRRC is readily derived as \citep{Schmidt2007b}
\begin{equation}\label{eq:alphaMBexp}
    \alpha_\mathrm{MB}(E_\mathrm{cm}) = R\frac{eqv_i}{(1-\beta_i\beta_e)I_in_eL\eta}.
\end{equation}
Here $eq$ is the charge of the primary ion, $v_i = c\beta_i$ and $v_e= c \beta_e$ are the ion and electron velocity, respectively, and $c$ denotes the speed of light in vacuum. The center-of-mass energy $E_\mathrm{cm}$ can be calculated from the laboratory ion and electron energies using  \citep{Schippers2000b}
\begin{equation}\label{eq:Ecm1}
    E_\mathrm{cm} = m_ic^2\left(1+\frac{m_e}{m_i}\right)\left[\sqrt{1+\frac{2m_e/m_i}{(1+m_e/m_i)^2}(\Gamma-1)}-1\right]
\end{equation}
with the electron and ion masses $m_e$ and $m_i$ and with
\begin{equation}\label{eq:Ecm2}
    \Gamma = \gamma_i\gamma_e-\sqrt{(\gamma_i^2-1)(\gamma_e^2-1)}\cos\theta
\end{equation}
where the Lorentz factors are $\gamma_i = 1/\sqrt{1-\beta_i^2}$, $\gamma_e = 1/\sqrt{1-\beta_e^2}$, and $\theta=0^\circ$ for a merged-beams arrangement with copropagating beams. In a storage-ring experiment $\beta_i$ is kept fixed and $E_\mathrm{cm}$ is varied by changing $\beta_e$ via the cathode voltage at the electron gun.

Further details of the various aspects of the experimental and data reduction procedures at the Heidelberg heavy-ion storage ring TSR have been discussed in depth in Refs.\ \citep{Kilgus1992,Pastuszka1996,Schippers2000b,Schippers2001c,Schippers2004c,Wolf2006c,Lestinsky2008a,
Schmidt2008a,Lestinsky2009}. The systematic experimental uncertainty of the measured MBRRC is typically 20\%--25\% at a 90\% confidence level.

As an example  Figure \ref{fig:Ni25} shows results for the recombination of Ni$^{25+}$ ions \citep{Schippers2000b}. The spectrum consists of DR resonances at specific energies sitting on top of the monotonically decreasing continuous rate coefficient due to radiative recombination (RR). The resonances in the energy range 0--75~eV are associated with $2s\to2p$ ($\Delta N=0$) excitations of the lithium-like $1s^2\,2s$ core, with the capture of the initially free electron into a Rydberg level $n$. Neglecting interactions between the outer Rydberg electron and the inner core electrons, the positions of the resonances can be estimated by applying the Bohr formula for hydrogenic ions of charge $q$ ($q=25$ for Ni$^{25+}$), i.e.,
\begin{equation}\label{eq:Ryd}
    E_n \approx E_\infty-13.606\mathrm{~eV}\times\frac{q^2}{n^2}
\end{equation}
For $n\to\infty$ the two $1s^2\,2p_{1/2}\,nl$ and $1s^2\,2p_{3/2}\,nl$ Rydberg series of DR resonances converge to their series limits at $E_\infty=52.92$~eV and $E_\infty=74.96$~eV, respectively. These energies correspond to the $2s\to2p_{1/2}$ and $2s\to2p_{3/2}$ excitation energies \citep{Ralchenko2008}. In Figure \ref{fig:Ni25} individual Rydberg resonances are resolved up to $n\approx 30$. The higher-$n$ resonances
are immersed in one broad structure due to the finite experimental electron energy-spread.

\section{Derivation and parameterization of DR plasma rate coefficients}\label{sec:plasma}

\ifprep \FigNiPlasma{ttt} \else \TabNi \FigNiPlasma{ttt}\fi

After subtraction of the continuous RR ``background'' from the measured recombination spectrum the DR PRRC is derived by convoluting the DR MBRRC with a Maxwell-Boltzmann electron energy distribution. As detailed in \citep{Schippers2001c,Schippers2004c}, there are three issues that require special consideration: the experimental energy spread, the recombination rate enhancement at low energies, and field ionization of high Rydberg states in the storage-ring bending magnets.

The experimental energy spread $\Delta E_\mathrm{cm}$ influences the outcome of the convolution for resonances with resonance energies $E_\mathrm{cm} \leq  \Delta E_\mathrm{cm}$. This can be circumvented by extracting the DR resonance strengths, e.g., by a fit of individual DR resonances to the measured MBRRC at low energies \citep{Schippers2004c}.

An enhanced MBRRC is consistently observed in merged electron-ion beam experiments with atomic ions at very low energies below a few meV. There, the measured MBRRC exceeds the theoretical expectation by factors of typically 2--3. This excess rate coefficient is an artifact of the merged-beams technique \citep{Gwinner2000,Hoerndl2006a}, and hence it has to be subtracted from the measured MBRRC prior to the calculation of the PRRC.

Field ionization of the loosely bound high Rydberg electron in the recombined ions can result from the motional electric fields that the ions experience inside the storage-ring bending magnets \citep{Schippers2001c}. For example, in the Ni$^{25+}$ experiment, only DR involving capture into Rydberg levels with quantum numbers less than 150 contributed to the MBRRC. The missing DR resonance strength up to $n_\mathrm{max} = 1000$ (where the PRRC has converged) can be estimated from a theoretical calculation using, e.g., the AUTOSTRUCTURE code \citep{Badnell1986}. For high Rydberg quantum numbers this code reproduces the regular DR resonance structure close to the Rydberg series limits reasonably well
(Fig.\ \ref{fig:Ni25plasma}a) especially when slight ``manual adjustments" are made in these calculations to the core excitation energies that are relevant for
the DR resonance positions and the DR rate coefficient scale. The resulting theoretical rate coefficient was multiplied by a factor
1.2 to match the experimental result in the energy range 60--73 eV (Fig.\ \ref{fig:Ni25plasma}a). The deviation of this factor from unity is within the
experimental error margin.

Figure \ref{fig:Ni25plasma}b displays the Ni$^{25+}$ DR PRRC which has been derived from the measured MBRRC using the procedures described above. The systematic uncertainty is basically the  20\% uncertainty of the experimental MBRRC. In some cases additional uncertainties arise from the subtraction of the non-resonant RR ``background''  from the measured MBRRC, from unresolved DR resonances at very low energies, from the theoretical estimate of the unmeasured high Rydberg resonances (especially for low charge states), or in certain cases from the presence of a then usually unknown fraction of primary ions in extremely long living metastable states such as the $2s\,2p\;^3P_0$ and $3s\,3p\;^3P_0$ states in berylliumlike ions \citep{Savin2006a,Schippers2007a,Orban2008a} and magnesiumlike ions \citep{Lukic2007a}, respectively.

A convenient parameterization of the DR plasma rate coefficient is
\begin{equation}\label{eq:PPRCfit}
    \alpha^\mathrm{DR}(T) = T^{-3/2}\sum_ic_i\exp(-E_i/T)
\end{equation}
where the parameters $c_i$ and $E_i$ are determined from a fit of equation \ref{eq:PPRCfit} to the experimentally derived DR plasma rate coefficient. It should be noted that a given set of parameters can only be used in a limited temperature range $[T_\mathrm{min}, T_\mathrm{max}]$. Outside this range the fit may deviate strongly from the experimentally derived curve. In the temperature range $[T_\mathrm{min}, T_\mathrm{max}]$ the deviation from the experimentally derived curve is usually less than 1.5\% so that no significant additional uncertainty is introduced by the parameterization. Table \ref{tab:Ni25} lists the PRRC parameters for DR of Ni$^{25+}$.

In the subsequent tables the temperature ranges of validity are also given along with the parameters $c_i$ and $E_i$ as well as the experimental uncertainties of the derived absolute DR plasma rate coefficient.
We note that in some of the references cited below the parameters $c_i$ and $E_i$ were tabulated in units that are partly different from the units used here.

\ifprep\TabNi\fi

\section{Results for iron ions}\label{sec:iron}

Before listing the fit parameters (cf.\ Equation \ref{eq:PPRCfit}) for the experimentally derived iron DR plasma rate coefficients that are available to date, several selected individual experimental results are discussed exhibiting peculiar aspects of electron-ion recombination physics.

\subsection{Importance of fine-structure core excitations}

\ifprep\FigFeSeventeen{ttt}\else\FigFeSeventeen{bbb}\fi

One of our first experiments on the iron isonuclear sequence with fluorine-like Fe$^{17+}$ ions \cite{Savin1997} revealed the importance of an effect that had been neglected in many previous theoretical calculations, namely DR associated with a fine-structure core excitation. For the Fe$^{17+}$ $2s^2\,2p^5\;(^2P_{3/2})\to 2s^2\,2p^5\;(^2P_{1/2})$ transition the corresponding excitation energy is rather low \citep[12.7182~eV,][]{Ralchenko2008} for such a highly charged ion. The associated series of $2s^2\,2p^5\;(^2P_{1/2})\,nl$ DR resonances dominates the low-energy DR spectrum (Fig.\ \ref{fig:Fe17}) and consequently also the PRRC at temperatures below $\sim$12~eV \cite{Savin1997}. Previous theoretical DR calculations for this ion had been carried out using the non-relativistic $LS$ angular momentum coupling scheme \citep{Roszman1987a,Dasgupta1990} which cannot account for fine-structure effects, or deliberately had disregarded fine-structure excitations to keep the computations manageable \citep{Chen1988a}. The resulting theoretical PRRC deviates strongly from the experimentally derived rate coefficient by  up to orders of magnitude (depending on temperature) \cite{Savin1997}.

In addition to the Fe$^{17+}$ DR measurements new theoretical calculations were carried out which included fine-structure core excitations \citep{Savin1997,Savin1999}. These new theoretical PRRC do not exhibit such a striking disagreement with the experimental result
as the earlier theoretical calculations. The remaining discrepancies are on a 30\% level which is larger than the estimated 20\% experimental uncertainty. The origin of these discrepancies is unclear.

\subsection{Low-energy DR resonances}

In storage-ring recombination experiments, low energy DR resonance positions can be measured with extreme precision. For example, in a TSR experiment with lithiumlike Sc$^{18+}$ ions an experimental uncertainty of less than 5 ppm has been achieved for DR resonances located at energies below 100~meV \citep{Lestinsky2008a}. These measurements are sensitive to higher order QED contributions to the $2s_{1/2}\to2p_{3/2}$ excitation energy of the lithiumlike core.

\FigFeSeven{bbb}

In contrast to this remarkable experimental precision theoretical calculations of DR resonance positions bear uncertainties of up to a few eV depending on the complexity of the atomic structure of the ion. These uncertainties are especially influential close to zero electron-ion collision energies \cite{Badnell2007b} where resonances may wrongly be predicted to exist within the  continuum of the primary ion while in reality the corresponding doubly excited states are bound, or vice versa. Thus, relatively small uncertainties in theoretical DR resonance positions can lead to large uncertainties of low-temperature DR rate coefficients \cite{Schippers2004c}.

Figure \ref{fig:Fe7}a gives an example for the typical differences between measured and theoretically predicted low-energy DR resonances of complex ions. Significant discrepancies can be seen between theoretically predicted  and experimentally measured  DR resonance structures for Fe$^{7+}$.
Similar discrepancies have been found for other members of the iron isonuclear sequence. Because of the general $1/E$ dependence of electron-ion recombination cross sections low-energy resonances are often quite strong and, consequently, make significant contributions to the low-temperature PRRC, as can be seen, e.g., from Fig.\ \ref{fig:Fe7}b. The discrepancies between theoretical and experimentally derived Fe$^{7+}$ PRRC are especially large at low temperatures where Fe$^{7+}$ forms in PP. For high-temperature DR uncertainties of DR resonance positions are much less influential. Consequently, there is usually much better agreement between theoretical and experimental DR rate coefficients at higher temperatures than at lower temperatures. Accurate low-temperature DR data for complex ions, however, can be generated at present only  from storage-ring experiments.

\subsection{Significance of trielectronic recombination}

\FigFeTwentytwo{ttt}

Trielectronic recombination (TR) is similar to DR. The difference is that TR is associated with the simultaneous excitation of two core electrons by the incoming free electron. TR is a particularly strong recombination process for Be-like ions where it proceeds via $2s^2\to2p^2$ core double-excitations. The first experimental observation of TR was made in an experiment at the TSR storage-ring with Be-like Cl$^{13+}$ ions \citep{Schnell2003b}. A detailed comparison with theoretical calculations revealed that, depending on plasma temperature,  for this particular ion TR contributes by up to 40\% to the PRRC. In subsequent storage-ring recombination experiments strong TR resonances have been found also for other members of the Be-like isoelectronic sequence \citep{Schnell2003a,Fogle2005a,Schippers2007b,Orban2008a}.  Recently, TR has also been observed in an electron-ion recombination experiment with carbonlike Kr$^{30+}$ ions using an electron-beam ion-trap \cite{Beilmann2009}.

In the measured Fe$^{22+}$ MBRRC (Figure \ref{fig:Fe22}) we can unambiguously identify only those few TR resonances which do not fit into the more regular pattern of DR resonances associated with $2s^2\to2s\,2p$ core single excitations. In particular, the first three members of the $2p^2\,{^1\!}D_2\;nl$ series of TR resonances with $n=7,8,9$ can be discerned. The TR resonance strengths decrease significantly beyond the various $2s\,2p\;^3P$ DR series limits where additional strong $2p^2\,nl \to 2s\,2p$ autoionization channels become available for the decay of the $2p^2\,nl$ intermediate states \citep{Schnell2003b}.

By analogy, TR resonances may also be expected for Mg-like ions where $3s^2\to 3p^2$ core double-excitations can occur. Unfortunately, the resonance structure of the MBRRC of Mg-like Fe$^{14+}$ is too rich to allow for a clear identification of TR resonances \citep{Lukic2007a}. It should be noted that TR is regularly included in state-of-the-art DR calculations \citep{Colgan2003a,Gu2003b,Gu2004a,Altun2007a}.

\subsection{DR associated with $\mathbf{\Delta N \geq 1}$  core excitations}

\FigFeEighteen{ttt}

The energy range that can be accessed in merged-beams recombination experiments at storage rings is practically unlimited. Even DR of the heaviest ``naturally'' available H-like ion, namely U$^{91+}$, was studied at the ESR storage ring in Darmstadt, Germany \citep{Brandau2006a}. Using a stochastically cooled $^{238}$U$^{91+}$ ion beam, KLL-DR resonance structures were observed at energies around 70~keV with an experimental energy spread corresponding to only a few times the natural linewidth. Storage ring experiments are thus well suited for also providing absolute rate coefficients for high-temperature $\Delta N\geq 1$ DR which is a significant cooling process in collisionally ionized plasmas \citep{Kallman2007a}.

Figure \ref{fig:Fe18}a shows the measured Fe$^{18+}$ MBRRC in the energy region of the $1s^2\,2s^2\,2p^3\,3\ell n\ell'$ $\Delta N=1$ DR resonances. Resonance groups associated with different $n$ of the outermost electron in the doubly excited intermediate states can be distinguished. The series limit $n\to\infty$ occurs at about 925~eV. In the temperature range where Fe$^{18+}$ forms in CP  the contribution of these $\Delta N=1$ DR resonances to the total Fe$^{18+}$ DR rate coefficient in a plasma is  up to an order of magnitude larger than the contribution by $\Delta N=0$ DR (Fig.\ \ref{fig:Fe18}b). On the other hand, $\Delta N=1$ DR is insignificant in the temperature range where Fe$^{18+}$ forms in a PP.

Our DR measurements with iron ions have so far been concentrating on $\Delta N=0$ DR and are therefore most relevant for astrophysical modeling of PP. Additionally, $\Delta N=1$ DR has been measured for some ions as detailed in Tab.\ \ref{tab:FeMshell} and Tab.\ \ref{tab:FeLshell}. Generally, modern theoretical calculations are in good accord with our experimental $\Delta N=1$ DR results.

\subsection{Experimentally derived DR plasma rate coefficients for Fe$^\mathbf{q+}$ ($\mathbf{q}$=7--10, 13--22)}

The experimentally derived DR plasma rate coefficients can be retrieved by using Eq.\ \ref{eq:PPRCfit} and the tabulated parameters $c_i$ and $E_i$ (listed in Tabs.\ \ref{tab:FeMshell} and \ref{tab:FeLshell}, see also Figs.\ \ref{fig:FeMshell} and \ref{fig:FeLshell}). Table \ref{tab:FeMshell} lists the parameters for the M-shell ions Fe$^{7+}$ \citep{Schmidt2008a}, Fe$^{8+}$ \citep{Schmidt2008a}, Fe$^{9+}$ \citep{Lestinsky2009}, Fe$^{10+}$ \citep{Lestinsky2009}, Fe$^{13+}$ \citep{Schmidt2006a}, Fe$^{14+}$ \citep{Lukic2007a}, and Fe$^{15+}$ \citep{Linkemann1995c,Mueller1999c} and Tab.\ \ref{tab:FeLshell} lists the parameters for the L-shell ions Fe$^{16+}$ \citep{Schmidt2009a,Zatsarinny2004b}, Fe$^{17+}$ \citep{Savin1999}, Fe$^{18+}$ \citep{Savin2002c}, Fe$^{19+}$ \citep{Savin2002a}, Fe$^{20+}$ \citep{Savin2003a}, Fe$^{21+}$ \citep{Savin2003a}, and Fe$^{22+}$ \citep{Savin2006a}.

The published experimentally derived Fe$^{13+}$ PRRC \citep{Schmidt2006a} comprises both DR and RR. Here the Fe$^{13+}$ PRRC is presented (Fig.\ \ref{fig:FeMshell}, Tab.\ \ref{tab:FeMshell}) without
the RR contribution. This Fe$^{13+}$ DR PRRC has been obtained by subtraction of Badnell's \citep{Badnell2006b} theoretical Fe$^{13+}$ RR plasma rate coefficient from the experimental DR+RR plasma rate coefficient \citep{Schmidt2006a}. The theoretical RR contribution \citep{Badnell2006b} to the total PRRC \citep{Schmidt2006a} is less than 5\% for $120\mathrm{~K}=T_\mathrm{min}  \leq T \leq 3.5\times10^6\mathrm{~K}$ and rises to 11.3\% at $T=T_\mathrm{max}=1.8\times10^7$~K. Thus, the additional uncertainty of the Fe$^{13+}$ DR PRRC due to the RR subtraction is assumed to be negligible.

The experimentally derived DR  plasma rate coefficient for DR of neon-like Fe$^{16+}$ \citep{Schmidt2009a} is in excellent agreement with the theoretical results of \citep{Zatsarinny2004b}. Therefore, no fit has been made to the experimental data and the theoretical fit parameters for Fe$^{16+}$ are given in Table \ref{tab:FeMshell}.

\ifprep
\TabFeMshell
\TabFeLshell
\clearpage
\fi

\FigFeMshell{ttt}

\FigFeLshell{ttt}

\section{Conclusions and outlook}

Our experimentally derived rate coefficients for dielectronic recombination of iron ions are particulary important for photoionized plasmas where highly charged ions form at relatively low temperatures and where storage-ring recombination experiments are presently the only source for reliable DR data.
However, the experimental resources are such that providing a DR data base for all astrophysically relevant ions is certainly prohibitive. We therefore hope that our results will be considered as valuable benchmarks guiding the future development of the theoretical methods. To this end, we plan to continue our DR measurements and to fill in the still missing gaps in the iron isonuclear sequence. Our results summarized in figures \ref{fig:FeMshell} and \ref{fig:FeLshell} clearly show that the derivation of plasma rate coefficients by interpolation across isonuclear sequences of ions is not an  appropriate approach.

\begin{acknowledgments}
The results could not have been obtained without the excellent support of the  accelerator and storage-ring crews of the Max-Planck-Institute for Nuclear Physics.  We are grateful to our colleagues Z. Altun, N. R. Badnell, E. Behar, J. Colgan, M. H. Chen, T. W. Gorczyca, M. F. Gu,  D. A. Liedahl, S. Loch, M. S. Pindzola, and O. Zatsarinny for theoretical support and  likewise to our experimental collaborators T. Bartsch, D. Bernhardt, M. Beutelspacher, C. Brandau, S. B\"{o}hm, A. Frank, M. Grieser,  G. Gwinner, D. Habs, A. Hoffknecht, J. Hoffmann, G. Hofmann,  S. M. Kahn, J. Kenntner, S. Kieslich,  A. Liedtke, J. Linkemann, D. V. Luki\'c, D. A. Orlov, R. A. Phaneuf, R. Repnow, G. Saathoff, A. A. Saghiri, E. Salzborn, D. Schwalm, M. Schmitt, M. Schnell, W. Spies, F. Sprenger, G. Wissler, O. Uwira, D. Yu, P. A. Z\'{a}vodsky, and S.-G. Zhou for their contributions to the research described in this review.
\end{acknowledgments}

\ifprep
\else
    \TabFeMshell
    \TabFeLshell
\fi

\clearpage


\end{document}